\documentclass{article}
\usepackage{arxiv}
\usepackage{amsmath}
\usepackage{amssymb}
\begin{document}
\renewcommand{\k}{{\vec k}}
\newcommand{\kap}{{\vec \kappa}}
\newcommand{\p}{{\vec p}}
\newcommand{\q}{{\vec q}}
\newcommand{\kp}{{\vec k'}}
\newcommand{\x}{{\vec x}}
\newcommand{\ap}{{\hat a}^\dagger}
\newcommand{\alp}{{\hat \alpha}^\dagger}
\renewcommand{\P}{\hat P}
\newcommand{\ks}{{\vec k \sigma}}
\newcommand{\ksp}{{\vec k' \sigma'}}
\renewcommand{\r}{{\vec r}}
\newcommand{\be}{\begin{equation}}
\newcommand{\ee}{\end{equation}}
\newcommand{\bea}{\begin{eqnarray}}
\newcommand{\eea}{\end{eqnarray}}

\title{What condensed matter physics and statistical physics teach us about the limits of unitary time evolution}
%
%\titlerunning{limits of unitary time evolution}  % abbreviated title (for running head)
\author{Barbara Drossel\\
%
%\authorrunning{Barbara Drossel}   % abbreviated author list (for running head)
Institute of condensed matter physics\\
Technische Universit\"at Darmstadt\\
 Hochschulstr. 6, 64289 Darmstadt, Germany}
%\email{barbara.drossel@physik.tu-darmstadt.de}
\maketitle              % typeset the title of the contribution
% \index{Ekeland, Ivar} % entries for the author index
% \index{Temam, Roger}  % of the whole volume
% \index{Dean, Jeffrey}

\begin{abstract}        % give a summary of your paper

The Schr\"odinger equation for a macroscopic number of particles is linear in the wave function, deterministic, and invariant under time reversal. In contrast, the
concepts used and calculations done in statistical physics and condensed matter physics involve stochasticity, nonlinearities, irreversibility, top-down effects,
and elements from classical physics.  This paper analyzes several methods used in condensed matter physics and statistical physics and explains how they are in fundamental ways incompatible with the above properties of the Schr\"odinger equation. The problems posed by reconciling these approaches to unitary quantum mechanics are of a similar type as the quantum
measurement problem. This paper therefore argues that rather than aiming at reconciling these contrasts one should use them to identify the limits of quantum mechanics. The thermal wave length and thermal time indicate where these limits are for (quasi-)particles that constitute the thermal degrees of freedom. 
%he abstract should summarize the contents of the paper
%using at least 70 and at most 150 words. It will be set in 9-point
%font size and be inset 1.0 cm from the right and left margins.
%There will be two blank lines before and after the Abstract.
%                         please supply keywords within your abstract
\end{abstract}
\section{Introduction}

Imagine a gas of $10^{23}$ atoms at a finite temperature $T$. On the most microscopic level, how do you envision the gas? Is it a set of balls that follow classical trajectories and have elastic collisions with each other? Or a set of wave packets that are so small at high temperatures that the particles resemble hard objects? Or is it a set of plane waves, such as are used in the calculation of the canonical partition function of the ideal gas? Or is it an entangled mess where the wave functions of all particles are present all over the volume and have a high degree of entanglement? 

This question is closely related to that of the quantum-classical transition and the measurement process: The Schr\"odinger equation leads to an entanglement of all particles that have interacted with each other. However, the classical picture is that of localized particles that resemble small balls more than  spatially broadly spread wave functions. In condensed matter physics and statistical physics, we are dealing with systems of $10^{23}$ particles at finite temperatures, and therefore we encounter the question of the quantum-classical transition at every corner. Solids have a crystal structure with localized atoms; conduction electrons scatter locally from defects; biomolecules have a complex shape with their atoms having well-defined positions. Usually, however, this question of the quantum-classical transition is not made explicit in condensed matter physics. The student who learns the material simple wonders why the methods employed in condensed matter theory and statistical physics deviate so much from a Schr\"odinger equation of many particles and appear over and again incompatible with it.

In this paper, I want to address this topic and expose the many different ways in which condensed matter physics and statistical physics differ from a many-particle Schr\"odinger equation. This will teach us a lot about the nature and scale of the transition from the quantum to the classical behavior. This means that condensed matter physics and statistical physics should be used as key sciences when trying to solve the puzzles posed by the interpretation of quantum mechanics.

Let us start with the Schr\"odinger equation: 
The Schr\"odinger equation is viewed by many as the 'Theory of Everything' \cite{laughlin2000theory} for the nonrelativistic systems described by condensed matter theory and statistical physics. For a many-particle system, it has the general form
\begin{equation}
i \hbar \frac{\partial}{\partial t}\Psi(\vec r_1, \cdots, \vec r_N,t) = \hat H \Psi(\vec r_1, \cdots, \vec r_N,t) \label{fullschroedinger}
\end{equation}
with
\begin{equation}
\hat H= \sum_{\alpha=1}^N \frac{\hat p_\alpha^2}{2m_\alpha} + \frac 1 2\sum\limits_{\alpha,\beta,\alpha\neq\beta}V_{\alpha\beta}(\vec r_\alpha - \vec r_\beta) \, ,\label{fullhamiltonian}
\end{equation}
Here,  $\alpha$ and $\beta$ count the particles of the system, which are either the atoms or the electrons and nuclei or the ions and valence electrons, depending on the considered system. The potential  $V_{\alpha\beta}(\vec r_\alpha - \vec r_\beta)$ describes the interaction between particles  $\alpha$ and $\beta$. Given the initial state $\psi(\vec r_1, \cdots, \vec r_N,0)$, future and past states are given by the relation
\begin{equation}
\Psi(\vec r_1, \cdots, \vec r_N,t) = e^{-i\hat H t/\hbar}\, \Psi(\vec r_1, \cdots, \vec r_N,0). \label{timeevolution}
\end{equation}
if the interaction potentials are not explicitly time dependent. 
\\

The Schr\"odinger equation has the following three important properties:
\begin{enumerate}
    \item It is linear.  This means that the time evolution of a superposition of two wave functions is given by the superposition of the two time evolutions. It further means that all particles that have interacted in the past are entangled. If the initial state is not entangled, i.e., if it is a product state 
    $$ \Psi(\vec r_1, \cdots, \vec r_N,0) = \prod\limits_{\alpha=1}^N \phi_\alpha(\vec r_\alpha)\, ,$$
    then the state at a later time will be a superposition of product states
    $$\Psi(\vec r_1, \cdots, \vec r_N,t) = \sum\limits_{i_1,\dots,i_N}c_{i_1,\dots,i_N}\prod\limits_{\alpha=1}^N \phi_{i\alpha}(\vec r_\alpha) \, .  $$ A further consequence of the linearity of the Schr\"odinger equation is that initially localized wave packets disperse and become delocalized.
    \item It is deterministic. This means that the initial state, combined with the Hamiltonian determines the future time evolution.
    \item It is invariant under time reversal. The time-reversed Schr\"odinger equation is solved by the complex conjugate wave function, and this does not affect the observables. 
\end{enumerate}
However, the world that surrounds us has none of these three properties. It consists of localized objects with classical properties that are not superpositions of states with different eigenvalues of the observables. The time evolution of macroscopic systems has nonlinear and stochastic elements, as becomes manifest in chaotic systems. Furthermore, when running backwards a movie of any process occurring in the real world, the process looks completely unrealistic. In fact, all three properties of the Schr\"odinger equation are  already violated within quantum mechanics, namely by the measurement process, without which no description of quantum mechanics is complete. During a measurement, an irreversible, stochastic 'collapse' to one of the possible measurement outcomes happens. 

There are a number of interpretations of quantum mechanics that attempt to explain away this dichotomy between the Schr\"odinger equation and the measurement process, however, to many people those interpretations remain unsatisfactory. 
These interpretations consider the mathematical description according to the Schr\"odinger equation to be the adequate description of the time evolution of a nonrelativistic system, include
the many worlds \cite{everett1957relative},  relational \cite{rovelli1996relational}, consistent (or decoherent) histories \cite{griffiths1984consistent}, modal, epistemic, de Broglie-Bohm, and statistical \cite{Ballentine} interpretations. These interpretations differ in the ontological status that they ascribe to the wave function and in their explanation of the observed randomness. Thus, for instance, the consistent histories interpretation does  not consider the wave function as real but uses the formalism to calculate  possible sequences of events and their probabilities, while Everett \cite{everett1957relative} considers the wave function as real but claims that our consciousness takes a stochastic route through the various branchings that occur at each measurement event. Epistemic interpretations, such as the  relational interpretation \cite{rovelli1996relational,rovelli2018space} or QBism \cite{fuchs2010qbism} argue that the actual values of the physical variables of a system are only meaningful in relation to another system with which this system interacts, or that they represent states of knowledge rather than the objective reality.
Similarly, modal interpretations are not concerned with an objective description independent of observers, but only in consistent classical descriptions of systems that are on a microscopic level described by quantum mechanics \cite{hollowood2017decoherence}. The statistical interpretation \cite{Ballentine} suggests that wave functions do not describe single systems but ensembles of identically prepared systems. The de Broglie-Bohm pilot wave theory is different in nature as it postulates hidden variables, namely the particle position and its deterministic time evolution, which depends non-locally on the wave function. The stochasticity of the measurements is encoded in the initial value of the hidden variable. 

In contrast to these interpretations, I will in this paper pursue the idea that there are limits of validity to the Schr\"odinger equation and that this is the reason why macroscopic objects behave differently than suggested by the three properties of the Schr\"odinger equation listed above. This is a view held by a variety of scientists, for instance by the condensed-matter theorist and Nobel laurate Anthony Leggett \cite{leggett1992nature}. In fact, the Copenhagen Interpretation also presupposes such limits as it maintains that a classical terminology for measurement outcomes and a classical environment are necessary for describing quantum systems. However, it remains ambiguous with respect to the ontological status of the classical and quantum worlds and the scale at which the 'Heisenberg cut' between the quantum and classical description occurs. 

Limits of unitary time evolution are implemented explicitly in models for stochastic wave function collapse \cite{bassi2013,gisin2017collapse}. They include stochasticity, localization, and irreversibility in the description of the time evolution of quantum systems from the onset. According to these models, the Schr\"odinger equation is only approximately correct as it neglects this stochastic component. The attractiveness of this alternative description lies in the fact that spontaneous collapses are extremely rare in single-particle systems but frequent in macroscopic systems. The shortcoming is that it contains ad-hoc parameters that cannot be derived from other theories, which is due to the fact that the specific context in which the collapse occurs does not feature in the description.% This applies in a similar way to other theories that include irreversibility and stochasticity at a fundamental level \cite{cortes2014quantum,sorkin1990spacetime}.

The approach that appears the most promising to me is to include top-down effects on a quantum system from its context in addition to the usual bottom-up approach of describing macroscopic systems in terms of their parts and the interactions between them. This approach has been advanced by G.~Ellis \cite{ellis2012} and more recently by Ellis and myself \cite{drossel2018contextual}.  
Such an approach has due to its realism the advantage that it does not require sophisticated interpretations quantum mechanics in order to reconcile the properties of the Schr\"odinger equation with those of the classical world. Furthermore, it presumes that the existing descriptions of condensed-matter systems are the most appropriate ones, as they have proven to be empirically adequate. What is thus lacking is not a new interpretation or a new, possibly more fundamental or exact, theory, but a fresh look at the calculations and concepts actually used in systems that consist of a macroscopic number of particles and have a finite temperature. 

For this reason, I will in the following consider those fields of theoretical physics that are most appropriate for describing macroscopic systems. These are condensed matter theory and statistical physics. It is my goal to show that the methods used in these  two areas of physics reveal us a lot about the transition from quantum to classical behavior. As these theories were developed as empirically adequate descriptions of the phenomena observed in macroscopic systems, they implement implicitly the quantum-classical transition in various ways. In fact, these methods include elements from classical physics as well as from quantum physics, and they violate all three above-listed properties of the Schr\"odinger equation. 

In the next two sections, I will first analyse several methods used in condensed matter theory, and then turn to statistical mechanics and its fundamental concenpts of probabilities and entropies. 

Some of the thoughts and arguments presented here have been discussed in my previous publications \cite{drossel2017ten,drossel16quantumclassical}.

\section{Condensed matter theory}

Condensed matter theory deals with phenomena such as superconductivity, magnetism, quantum Hall effect, metal-insulator transition, Anderson localization, and many more. Text books and research papers on condensed matter  theory are full of quantum mechanical equations that capture the important features of these phenomena. However, these quantum-mechanical equations are never the 'Theory of Everything' \eqref{fullschroedinger}, but effective equations for only some of the degrees of freedom of the system. When setting up these equations, empirical knowledge about the phenomena to be described and a basic understanding of these phenomena in terms of a model are essential. This is beautifully summarized by A. Leggett in an article that he wrote at the occasion of the 90th birthday of Karl Popper \cite{leggett1992nature}: 
\begin{quote}
No significant advance in the theory of matter in bulk
has ever come about through derivation from microscopic principles. (...) I would confidently argue further that it is in principle
and forever impossible to carry out such a derivation. (...) The so-called derivations of the results of solid state
physics from microscopic principles alone are almost all bogus, if
'derivation' is meant to have anything like its usual sense. Consider as
elementary a principle as Ohm's law. As far as I know, no-one has ever
come even remotely within reach of deriving Ohm's law from microscopic
principles without a whole host of auxiliary assumptions ('physical
approximations'), which one almost certainly would not have thought of
making unless one knew in advance the result one wanted to get, (and
some of which may be regarded as essentially begging the question). This
situation is fairly typical: once you have reason to believe that a certain
kind of model or theory will actually work at the macroscopic or intermediate
level, then it is sometimes possible to show that you can 'derive'
it from microscopic theory, in the sense that you may be able to find the
auxiliary assumptions or approximations you have to make to lead to the
result you want. But you can practically never justify these auxiliary
assumptions, and the whole process is highly dangerous anyway: very often
you find that what you thought you had 'proved' comes unstuck
experimentally (for instance, you 'prove' Ohm's law quite generally only
to discover that superconductors don't obey it) and when you go back to
your proof you discover as often as not that you had implicitly slipped in
an assumption that begs the whole question. (...) 
Incidentally, as a psychological fact, it does occasionally happen that one
is led to a new model by a microscopic calculation. But in that case one
certainly doesn't believe the model because of the calculation: on the contrary,
in my experience at least one disbelieves or distrusts the calculation
unless and until one has a flash of insight and sees the result in terms of
a model.
I claim then that the important advances in macroscopic physics come
essentially in the construction of models at an intermediate or macroscopic
level, and that these are logically (and psychologically) independent of
microscopic physics. Examples of the kind of models I have in mind which
may be familiar to some readers include the Debye model of a crystalline
solid, the idea of a quasiparticle, the Ising or Heisenberg picture of a
magnetic material, the two-fluid model of liquid helium, London's
approach to superconductivity .... In some cases these models may be
plausibly represented as 'based on' microscopic physics, in the sense that
they can be described as making assumptions about microscopic entities
(e.g. 'the atoms are arranged in a regular lattice'), but in other cases (such
as the two-fluid model) they are independent even in this sense. What all
have in common is that they can serve as some kind of concrete picture,
or metaphor, which can guide our thinking about the subject in question.
And they guide it in their own right, and not because they are a sort of
crude shorthand for some underlying mathematics derived from 'basic
principles.' 
\end{quote}
In the following, I will highlight these characteristics of condensed matter theory by describing several widely used methods.

\subsection{Born-Oppenheimer approximation}
The Born-Oppenheimer approximation is the basis of quantum chemistry and solid state physics. It separates the quantum mechanical equation of the electrons from that of the ions. 

The following derivation follows closely the textbook 'Advanced Quantum Mechanics' by F. Schwabl \cite{schwabl2005advanced}. The Born-Oppenheimer approximation starts from the  Hamilton Operator for all electrons and ions,
\be
\hat H = T_e + T_I+V_{ee}+V_{eI}+V_{II}
\ee
with the kinetic energy  $$T_e=\sum_i\frac{\hat p_i^2}{2m}\, \text{ and } T_I=\sum_K \frac{\hat P_K^2}{2M}$$ of electrons and ions, respectively, and with the three contributions to electrostatic energy between electrons and electrons, electrons and ions, and ions and ions. We focus on the time-independent Schr\"odinger equation
$$ H\Psi(\vec x,\vec X)=E\Psi(\vec x,\vec X)\, , $$
as this method is usually used to find the ground state. We make the ansatz
\be
\Psi(\vec x,\vec X)=\psi(\vec x|\vec X)\Phi(\vec X)\, ,
\ee
which is a product of a wave function $\Phi(\vec X)$ for the ions and a wave function $\psi(\vec x|\vec X)$ of the electrons (with given ion positions). The vector $\vec X$ denotes the positions $(\vec X_1, \dots, \vec X_N)$ of the ions, and the vector $\vec x$ the positions  $(\vec x_1, \dots, \vec x_N)$ of the electrons. The wave function of the electrons for given positions $\vec X$ of the ions is determined from the reduced Schr\"odinger equation 
\be
(T_e+V_{ee}+V_{eI}+V_{II}) \psi(\vec x|\vec X)=\epsilon(\vec X)\psi(\vec x|\vec X)\, .
\ee
The energy eigenvalue has the contributions $\epsilon(\vec X) = V_{II}(\vec X) + E^{el}(\vec X)$, which are the interaction energy of the ions and the energy eigenvalue of the electrons in the potential of the ions. In order to obtain the ion wave function  $\Phi(\vec X)$, the reduced Schr\"odinger equation is inserted into the full Schr\"odinger equation, giving 
\bea
&& \psi(\vec x|\vec X)(T_I+\epsilon(\vec X))\Phi(\vec X) = \psi(\vec x|\vec X)E\Phi(\vec X)\nonumber \\
&& \quad -\sum_K\frac{-\hbar^2}{2M}\left[\Phi(\vec X)\vec \nabla_X^2\psi(\vec x|\vec X)+2\vec \nabla_X\Phi(\vec X)\cdot \vec \nabla_X\psi(\vec x|\vec X)\right]\, .
\eea
Multiplication with  $\psi^*(\vec x|\vec X)$ and integration over $\vec x$ and neglection of two terms gives the Born-Oppenheimer equation 
\be
(T_I+\epsilon(\vec X))\Phi(\vec X) = E\Phi(\vec X)\, .
\ee
Here, the energy contribution $\epsilon(\vec X)$ due to the electrons is responsible for an effective attractive interaction between the ions. The equilibrium positions of the ions are obtained by minimizing $\epsilon(\vec X)$. When these minima are sufficiently pronounced, the ions are well localized. The second derivatives of $\epsilon(\vec X)$ at the positions of the minima determine the oscillation frequencies of the ions and yield the vibrational spectra of molecules or the phonon spectrum of crystals. The two neglected terms can be argued to be smaller by a factor $m/M$ than the leading terms, with $M$ being the mass of the ions and $m$ the mass of the electrons. The intuitive justification given for this procedure is that the electrons move much faster than the ions and thus are at all times in equilibrium with the present position of the ions. This is also called the adiabatic approximation. Philosophers of chemistry point out the the Born-Oppenheimer approximation is in fact a mixture of classical and quantum mechanics
\cite{primas2013chemistry,chibbaro2014reductionism,matyus2018pre}.
By making a product ansatz, the entanglement between electrons and ions is ignored. By presuming localized ions, spatial superpositions of different ion positions are ruled out. This is the basis for molecules assuming specific shapes that break the symmetry of the underlying full Schr\"odinger equation. The breaking of a symmetry is something similar to the measurement process, since in both cases only one of several possible options becomes realized. In fact, it is well known that the ground state of a wave function cannot break a symmetry of the Schr\"odinger equation if the degree of freedom that shows this symmetry occurs explicitly in the Hamilton operator. 

To summarize, the Born Oppenheimer approximation violates the linear superposition principle of the Schr\"odinger equation. It mixes classical with quantum features in order to obtain molecular configurations and crystal structures. Most subfields of condensed matter theory take such structures for granted, for instance when they  deal with the properties of electrons in crystals.

\subsection{Nonlinear wave equations}

The linearity of the Schr\"odinger equation is essential for obtaining superpositions and entanglement. Despite the importance of this feature, condensed matter physicists use various equations that are nonlinear in the wave function. The best-known examples are the Hartree and Hartree-Fock equations. Both are used to approximate the ground state of a many-electron system by a variational calculation that expresses the many-particle wave function in terms of a product of one-particle wave functions, which is additionally antisymmetrized in the Hartree-Fock approach. In the Hartree approach the ansatz for the spatial part of the wave function is therefore    
\be
\psi(\r_1,\dots,\r_N)=\phi_1(\r_1) \phi_2(\r_2)\dots \phi_N(\r_N)\, ,\label{phiprod}
\ee
Minimizing  the expectation value of the Hamilton operator with respect to the one-particle wave functions with the constraint that all wave functions shall be normalized gives 
\be
\left(-\frac{\hbar^2}{2m}\nabla^2 + U(\vec r) + \sum_{j\neq i}\int d^3 r' \frac{e^2}{|\vec r - \vec r\:'|}|\phi_j(\vec r\:')|^2\right)  \phi_i(\vec r) =  \epsilon_i\phi_i(\vec r) \, .\label{hartree}
\ee
This is a nonlinear equation in the one-particle wave functions that is to be solved self-consistently with the condition that the states of the different particles are orthogonal to each other. They have the form of mean-field equations  where each particle is subject to a classical electrostatical potential that is caused by the charge density of the other particles. 
Just as for the Born-Oppenheimer approximation, we see here that elements from classical physics enter the quantum mechanical calculation. An additional classical feature was already put into the system at the beginning of the calculation: the product ansatz rules out entanglement.

In the Hartree-Fock method, equations \eqref{hartree} obtain an additional term that comes from the antisymmetrization. A similar method is also used for calculating the ground state of interacting bosons, giving the Gross-Pitaevskii equation
\be
\left(\frac{-\hbar^2}{2m}\nabla^2 + U(\r) + g |\psi(\r)|^2\right) \psi(\r) = \epsilon \psi(\r)\, .
\ee
Here, the wave function $\psi(\r)$ is normalized such that $\langle \psi|\psi\rangle = N$, with $N$ being the particle number.
Since bosons all go into the same one-particle ground state, only one state occurs in this equation. Since the atoms of a Bose gas have a short-range interaction, the potential has been approximated as $V(\r-\r') = g \delta(\r-\r')$.
Here arises another remarkable feature: the interpretation of the wave function changes. The macroscopic wave function of the ground state of a bosonic system is not interpreted as a probability amplitude but as a partially classical object. In particular, the states of systems that contain $N$ and $N+1$ particles respectively are not assumed to be orthogonal to each other (as required by the quantum-mechanical Fock-space formalism) but to be parallel to each other and to differ only by a factor $\sqrt{(N+1)/N}$.  

Similar nonlinear and classical features occur in density-functional theory, which calculates the ground state charge density of the electrons of a molecule or crystal. In the Kohn-Sham approach \cite{kohn1965self}, the effective one-particle Schr\"odinger equation to be solved takes the form
\be
\left[\frac{-\hbar^2}{2m}\nabla^2 + V_{eff}(\vec r)\right]\phi_i(\vec r) = \epsilon_i \phi_i(\vec r)\, .
\ee
The effective potential
\be
V_{eff}(\vec r) = U_{ext}(\vec r)  + \int d^3 r\:' e^2\frac{ n(\vec r\:')}{|\vec r -\vec r\:'|} + V_{xc}(\vec r)
\ee
includes the external potential that comes from the nuclei, the long-range electrostatic repulsion of the electrons in the form of a Hartree term, and a short-range exchange-correlation potential that captures all effects that have been neglected in the other terms. This means that entanglement and antisymmetry of electrons are only taken into account on short length scales, while the long-range part of the interaction is assumed to be classical. 
Even though the one-particle wave functions are interpreted only as an auxiliary quantity to calculate the ground-state charge density and not as the real states of the electrons, this does not change the fact that density functional theoretical calculations use successfully nonlinearities and classical ingredients. It considers linear superposition and entanglement only on short distances.

\subsection{Molecular Dynamics simulations}

Molecular dynamics (MD) simulations are successfully used to characterize the structure and dynamics of systems that consist of many atoms at a finite temperature \cite{tuckerman2000understanding,marx2000ab}. They include quantum mechanical features only in a very limited way, if at all. In purely classical simulations, molecules are represented as a collection of point masses and charges that obey Newton's laws.  When the formation and breaking of bonds, the polarization of atoms or molecules, or excited states shall be taken into account, the quantum mechanical properties of the electrons must be considered, employing ab initio MD simulations. For given positions of the nuclei, the electronic structure of atoms and molecules is calculated using Density Functional Theory. The motion of the nuclei is then calculated classically based on the force fields resulting from the electronic structure, and the electronic structure in turn is recalculated based on the changed positions of the nuclei. When quantum mechanical properties of nuclei become important, for instance with proton transfer processes that involve tunnelling,  the Feynman path integral formalism of statistical
mechanics is used to describe the nuclei. Again, quantum effects are only included on short length scales, and the state of $N$ particles is calculated as a product, as if the particles were distinguishable and not entangled with each other. 

In plasma physics, and when describing electrons in a conductor, another version of MD is used, which is called wave packet molecular dynamics \cite{grabowski2014review}. This type of simulation assumes that the particles are wave packets all the time. These wave packets are modeled by their center-of-mass coordinate and their width, and it is assumed that they have Gaussian shape. The computer simulation therefore does not run the Schr\"odinger equation, but a classical equation where forces and velocities are obtained by their quantum mechanical averages, and this is justified by the Ehrenfest theorem.

In all these different methods for performing MD simulations, the atoms are localized: they are points for classical MD simulations, they have the extension of the electronic shell for ab initio simulations, and the extension of the thermal wavelength for path integral ab initio simulations. In none of these approaches are the atoms or molecules entangled with each other, even though the time evolution of the system according to the full Schr\"odinger equation for all particles would yield such an entanglement. The success of MD calculations shows that they are adequate  to capture the phenomena observed in these systems.

\subsection{Linear Response theory}

Linear response theory is a conceptually very interesting theory as it breaks time reversal invariance and relies on the concept of causality in its old-fashioned sense: first the cause, then the effect. It
is an important tool for connecting theory with experiments as it is employed to calculate the response of a system to an applied field $F$, such as an electrical or magnetic field or a pressure. If $\hat B$ is the observable to which the field couples and $\hat A$ the observable that is evaluated, the response of the expectation value of this observable to the field is obtained in the Heisenberg picture as \cite{schwabl2005advanced}
\be
\langle \hat A(t)\rangle - \langle \hat A \rangle_0 = \int_{-\infty}^\infty dt' \, \chi_{AB}(t-t')F(t')\label{DeltaA}
\ee
with the susceptibility
\be
\chi_{AB}(t-t') = \frac i \hbar \theta(t-t') \langle[\hat A(t),\hat B(t')]\rangle_0\, .\label{chiAB}
\ee
Here, the expectation value $\langle \dots\rangle_0$ is evaluated in thermodynamic equilibrium using the canonical or grand canonical ensemble. The step function $\theta(t-t')$ implements causality such that the effect of the field at time $t'$ is only felt at times $t > t'$. The derivation of these expressions presumes implicitly that the system is in thermal equilibrium when no field is applied and that it returns to this equilibrium after the field ceases to act. These assumptions are included as an additional feature, and they cannot be justified by the many-particle Schr\"odinger equation. They are in fact a version of the  second law of thermodynamics. 

This leads us to the next section, which discusses in more detail how statistical physics is done and how it also violates the properties of the many-particle Schr\"odinger equations.

\subsection{Interim discussion}

The examples of condensed matter calculations given so far are only a small glimpse into the methods and concepts used in this field. In fact, one could write an entire book that analyses the methods and calculations used in condensed matter theory. Over and again, one can see that elements foreign to a many-particle Schr\"odinger equation are introduced. In fact, all three properties of the Schr\"odinger equation are violated in condensed matter theory: linearity, determinism, and time reversal invariance. Linear superposition and entanglement are taken into account only over short distances, as we have seen in the previous subsections. Determinism is broken wherever transition probabilities are used. This becomes most evident in calculations of (solid-state) quantum field theory and scattering theory, both of which were not mentioned explicitly so far. Transition probabilities are also implicit in statistical physics, which was mentioned in the context of linear response theory. Time reversal invariance is broken whenever response functions are used. In addition to the example mentioned above, this happens also in scattering theory, where the incoming wave knows nothing about the scattering potential, but the outgoing wave does, and in quantum field theory, where propagators are the microscopic equivalent of response functions. 

In the next section, we will discuss how statistical mechanics also violates all three properties of the Schr\"odinger equation.

\section{Statistical Mechanics}

\subsection{Statistical independence}
The textbook of Landau and Lifschitz on Statistical Physics \cite{landau2013course} begins with the assumption that subsystems of a larger system are statistically independent. This means that the probability of a subsystem to be in a given microstate is independent of the microstate of neighboring subsystems. It is only determined by the macrostate, i.e., by the state variables. This basic assumption is a necessary requirement for observables to be sum variables and for the extensivity of thermodynamic systems. The standard derivation of the Boltzmann probabilities in the canonical ensemble, as done for instance by Schwabl \cite{schwabl2006statistical} or Kittel and Kroemer \cite{kittel1980thermal}, relies on this statistical independence, as it makes a product ansatz for the probabilities of  a subsystem and the remaining system to be in their different possible microstates. 

The assumption of statistical independence is logically incompatible with entanglement, as entanglement implies that specific states of one subsystem are correlated with specific states of the other subsystem. Statistical physics is therefore incompatible with a unitary time evolution of a many-particle system as it always leads to entangled states. We have already seen in the previous section that calculations in condensed matter physics often include entanglement only on short length scales. In fact, whenever a product ansatz is used in quantum mechanical calculations, statistical independence is implicitly presumed, and entanglements that should result from past interactions are neglected. A product ansatz is often used when dealing with quantum systems and their environment, for instance in models of the measurement process \cite{zurek2003decoherence}, or in the derivation of the Lindblad equation \cite{breuer2002theory}.  

Such a product ansatz is successful not because it is justified by the Schr\"odinger equation, but because it is in agreement with the observed properties of the respective systems. It thus appears that cutting entanglement beyond a certain distance is an appropriate procedure if one wants to capture what goes on in finite-temperature systems.

\subsection{Probabilities}

Probabilities are a fundamental, irreducible concept in statistical mechanics. They are introduced at the beginning of the considerations and derivations performed in statistical mechanics. Often, all the equilibrium properties in statistical mechanics are derived from the basic axiom that in an isolated system in equilibrium all microstates that are compatible with the macroscopic state variables (such as energy, volume, particle number) occur with the same probability, see, e.g., \cite{kittel1980thermal}. 

The concept of probabilities is foreign to a deterministic theory, such as the many-particle Schr\"odinger equation. Although several authors try to 'derive' the axiom of equal probabilities from a deterministic theory \cite{eisert2015quantum} or to ascribe it to our subjective ignorance of the precise microscopic state of the system \cite{jaynes1957information,landau2013course}, they cannot do this consistently, but need additional assumptions, which amounts to encoding all the  stochasticity of the trajectory in the initial state \cite{gisin2018indeterminism,drossel2017ten}. Then the basic axiom takes the following form: Among all the possible states that are compatible with a certain limited precision of the initial state, each state has the same probability to be  the actual initial state. Only then can one claim that the vast majority of these initial states leads to a homogeneous distribution in phase space and therefore to equal probabilities of the microstates in equilibrium. The mentioned problem of introducing probabilities in a deterministic theory exists already for classical mechanics, but becomes much worse in quantum mechanics. The reason is that the deterministic time evolution of a many-particle Schr\"odinger equation leads to a superposition of all possible classical outcomes (i.e., particle configurations), even when the above arguments about the limited precision of the initial state are applied and decoherence makes the density matrix diagonal with a very high probability. For instance, when a gas of atoms is considered, one can make plausible arguments that the density matrix of a subsystem loses correlations in space and is in position representation nonzero only along a stripe around the diagonal \cite{popescu2006entanglement}. Such a density matrix can be interpreted as a superposition of different classical states of the gas, in each of which the atoms are localized to a large degree. This is a situation similar to the measurement problem: decoherence might give a diagonal density matrix, but for a single run of a time evolution it does not give  only one classical outcome but all of them at once. If physics shall agree with our observation of only one outcome, this is not satisfactory. 

It thus seems that the probabilities used in statistical mechanics point to limits of validity of a deterministic time evolution. In fact, the very concept of entropy suggests that the state of a finite-temperature system is specified only by a limited number of bits. In an isolated system, we have in equilibrium $S=k_B \ln \Omega$, with $\Omega$ being the number of different microstates. In contrast, a wave function requires an infinite number of bits to be fully specified, as it is a complex-valued function. 

If states are specified only by a limited number of bits, this leads directly to a stochastic time evolution \cite{gisin2018indeterminism,grangier2018quantum,drossel2017ten} since the initial state does not yet unequivocally specify the full future time evolution. This in turn means that there must be stochastic transitions or 'collapses' (which could be continuous in time) that specify which of the possible time evolutions is actually taken. 

Stochasticity of the time evolution in turn leads to irreversibility: This becomes most evident by looking at equations for the time evolution of probability distributions or ensembles, such as the Boltzmann equation (for a gas of particles that make elastic collisions), the Fokker-Planck equation (for a stochastic reaction system or other classical stochastic process), or the Lindblad equation. All of them converge to a unique stationary distribution under suitable conditions, and this means that they can be associated with a Liouville function that never increases. In particular, the initial state is forgotten after some time and cannot be deduced any more from the later state. 

%..... hier fehlt noch was: extensives System, viele Teilchen /Raumelemente

So we see that the very concept of probabilities violates all three properties of the Schr\"odinger equation: it limits superpositions, determinism, and reversibility. Furthermore, these features of statistical physics reflect what happens in a measurement process. In their textbook on statistical mechanics \cite{landau2013course}, Landau and Lifschitz suggest that the irreversibility arising from statistical physics and that of the measurement process might in fact have the same origin. Indeed, the methods developed in the meantime in the fields of the foundations of statistical mechanics and in quantum foundations are very similar. These are based on decoherence theory and the theory of open quantum systems and acknowledge the importance of heat baths with a macroscopic number of degrees of freedom. However, all these methods introduce features that are incompatible with unitary time evolution, as will be discussed further in the next subsection.

\subsection{Maximum entropy principle}

The maximum entropy principle is more general than the principle that in an isolated system in equilibrium all microstates have the same probability. It applies also to systems in contact with a heat bath or a particle reservoir. The maximum entropy principle \cite{jaynes1957information} states  that in equilibrium the probabilities $p_n$ of the microstates are such that entropy $S=-k_B\sum_n p_n \ln p_n$ is maximized, given the relevant constraints. 
For a system in contact with a heat bath this constraint is a fixed mean energy, which is imposed on the system by the temperature of the environment. Maximizing entropy for a fixed mean energy and denoting the Lagrange multiplier that is introduced to fix the mean energy as $1/k_BT$, one obtains the probability to be in the microstate $n$ as  $$p_n = \frac{e^{-E_n/k_BT}}Z\, ,$$ with $E_n$ being the energy of microstate $n$, and $Z=\sum_n e^{-E_n/k_BT}$ the partition function. Similarly, imposing a mean energy and a mean particle number on the system leads to the grand canonical probabilities $$p_n = \frac{e^{-(E_n-N_n\mu)/k_BT}}{Z_{GK}}\, ,$$ with $\mu$ being the chemical potential and $N_n$ the particle number of microstate $n$ and $Z_{GK}= \sum_ne^{-(E_n-N_n\mu)/k_BT}$. For an isolated system, there are no constraints, and all probabilities are equal. 

The principle of maximum entropy means essentially that a system in equilibrium is completely indifferent with respect to the microstate in which it should be. Often, this principle is interpreted as being due to our ignorance of the precise microstate, but why should the system care about our ignorance when choosing its microstates? If there was anything else unknown to us determining the state, the principle of maximum entropy would not be empirically adequate, and the results derived from it would not agree with observation. But equilibrium statistical mechanics, which is based on the above-given probabilities, yields a wealth of results that agree with the experiment, for instance the behavior of specific heat at high and low temperatures, the temperature of Bose-Einstein condensation, the spectrum of black-body radiation, and many more.

All this means  that a system in equilibrium is determined solely by the macroscopic state variables imposed on it from the outside, classical world. This is a clear-cut example of top-down causation \cite{ellistopdown}. Notice again the similarities to the measurement process: the context (i.e., the experimental setup) determines the possible measurement outcomes and their probabilities. Of course, both are also affected by the incoming wave function, but this is again similar to a statistical physics system: the possible energy eigenstates depend on intrinsic properties, such as the masses and interactions of the particles. 

The principle of maximum entropy is in fact applied regularly in the foundations of statistical mechanics and in open quantum systems, usually without mentioning this explicitly. Most of these 'derivations' use concepts from decoherence theory \cite{gogolin2016equilibration,reimann2013quantum,popescu2006entanglement}. They invoke 'typical' environmental states, which are random ones among all those that are compatible with our knowledge of the system, as mentioned above. This means that these environmental states are determined by nothing but the classical variables that we know of or can control. The 'derivation' of the second law of thermodynamics by using the eigenstate thermalization hypothesis \cite{srednicki1994chaos,deutsch1991quantum} also employs this principle: the starting point is the plausible hypothesis that thermodynamic observables evaluated in an eigenstate of a macroscopic system show the properties of thermal equilibrium, such as an even distribution of density in space. Now, if a special initial condition is chosen, such as all atoms of a gas being in one corner of a box, this initial condition can be written as a superposition of eigenstates. This superposition is special in the sense that it is associated with an untypical density distribution.  But in order to argue that the system will evolve to an even density distribution, one must assume that the inititial superposition is not special with respect to the future time evolution, but that it is a random or 'typical' one among all those that represent the initial density distribution. Only then can one expect that the superposition of eigenstates does not lead to special or 'unlikely' future states.  

A similar analysis can be performed of the 'derivations' of the Lindblad equation from a quantum system coupled to its environment, with the total system being modelled quantum mechanically. Again, the environmental states must be assumed to be not correlated in a special way but to forget the past dynamics, if the Lindblad equation shall result in the end for the dynamics of the reduced density matrix of the quantum system. 

To summarize this subsection: the principle of maximum entropy is applied in various ways in the field of foundations of statistical mechanics and of open quantum systems. Whenever 'typical' states are invoked, this means essentially that the system (or environment) is determined by nothing else but the factors that are already taken into account.

\subsection{Thermal wave length and thermal time}

If we take seriously the principles of statistical independence, stochasticity, and maximum entropy on which statistical physics is based, we must conclude that unitary time evolution and linear superposition according to the Schr\"odinger equation are valid only under restricted conditions and are otherwise limited in length and time scale. In fact, there are many reasons why the most appropriate quantum mechanical description of particles and quasiparticles in finite-temperature systems is a wave packet of limited extension that remains a wave packet under time evolution: (i) The treatment of electrons in conductors as wave packets (ii) The treatment of particles in MD simulations as localized objects in space (iii) The requirement that in the classical limit of high temperature the atoms of a gas should become classical particles (iv) The treatment of the atoms of a gas as wave packets when it is argued that Bose-Einstein condensation sets in when these wave packets begin to overlap due to decreasing temperature or increasing density (v) The limited number of bits available for describing the state of a gas or other many-particle systems, as given by the entropy.

However, if wave packets  remain wave packets under time evolution, this time evolution cannot be exclusively determined by a Hamilton operator since this would lead to a spreading of the wave packet in space due to dispersion and scattering events. Therefore, an ongoing process of localization must occur that is not captured by the Schr\"odinger equation. Such localization processes are indeed implemented in  Lindblad equations for open quantum systems \cite{breuer2002theory} and stochastic collapse models \cite{bassi2013}.  

The width of thermal wave packets is given by the thermal wavelength. This is for instance presumed when the following quick derivation of the Bose-Einstein condensation temperature of an ideal Bose gas is given: Bose-Einstein condensation happens when the density of the gas becomes so large that the distance between atoms becomes of the order of the thermal wave length. This leads to 
$(V/N)^{1/3} = \lambda_{th} = h/\sqrt{2\pi m k_B T} $ and consequently $T = (N/V)^{2/3}h^2/(2\pi m k_B)$, which is apart from a numerical factor of the order 1 identical with the condensation temperature derived from the fully-fledged calculation. The thermal wave length emerges naturally in calculations of the canonical partition function of an ideal gas. Its order of magnitude can also be estimated without performing calculations from the equipartition theorem  $\frac 3 2 k_BT=E=h^2/(2m\lambda_{th}^2)$. The ratio $\lambda_{th} ^3/V$ is often called the \emph{quantum concentration} (see for instance the textbook by Kittel and Kroemer \cite{kittel1980thermal}).   When the density is so high that the wave packets overlap, bosons tend to go into the same quantum state. In the opposite case that the density is so small that the wave packets are not in contact for most of the time, the gas can be approximated as a classical ideal gas of well localized atoms. Similar arguments can be made for fermionic gases: When the density is so large that all the wave packets touch each other, it cannot be further increased due to the Pauli principle, leading to a Fermi temperature that is up to a constant factor identical with the Bose-Einstein condensation temperature. Again, in the limit of very low density, quantum effects become completely irrelevant, and the Fermi gas can be treated like a classical gas. For thermally equilibrated fermions as well as for bosons, the specific quantum mechanical effects become thus only important when the concentration is not small compared to the quantum concentration. 

For massless particles, such as photons and phonons, the thermal wavelength is $\pi^{3/2}\hbar c/k_BT$, with $c$ being the velocity of light and sound, respectively. Interestingly, the density of photons and phonons in thermal equlibrium adjusts itself such that the average distance between these particles is of the same order as this thermal wavelength. 

Closely associated with the thermal wavelength is a thermal time proportional to $\hbar/k_BT$. 
In the non-relativistic quantum theory of open systems, it sets the time scale beyond which the dynamics of the system shows the Markov property \cite{weiss2012quantum}. This time is of the order of the time that a wave packet with a kinetic energy of the order of $k_BT$ needs to cover the distance of the thermal wavelength. In field-theoretical treatments of condensed matter physics, the inverse of the thermal time occurs as the Matsubara frequency.

All the foregoing considerations suggest that the thermal wavelength and thermal time of the degrees of freedom that constitute the heat bath set the length and time scales for quantum coherence and linear superposition for all particles that are part of the heat bath and all those that are in equilibrium with it and  exchange energy with it. This view is to be distinguished from decoherence theory, which also gives a loss of quantum mechanical coherence beyond the thermal wavelength and thermal time. However, according to decoherence theory the quantum system and the heat bath are  described by a combined Schr\"odinger equation, and the loss of quantum coherence in the considered quantum system is due to an entanglement with the heat bath.  In contrast, the view advocated here is that there is no global wave function for the combination of quantum system and heat bath but that the dynamics of the heat bath is described by a (partially) stochastic time evolution of (quasi-)particle wave packets.

\section{Conclusions}

Let us come back to the finite-temperature gas mentioned in the Introduction.  We have seen that molecular dynamics simulations would treat such a gas quantum mechanically only on short length scales of the order of the thermal wavelength and classically beyond. Similarly, statistical mechanics requires the inclusion of quantum superposition only on distances of the order of the thermal wavelength. This means that the representation of the gas atoms as wave packets, which remain wave packets under time evolution, is the most appropriate one. For solids, the situation is similar: The lattice structure of a crystal presumes localized ions, and theories of electrical resistance all rely on localized conduction electrons. Very generally, statistical independence of subsystems is a starting assumption of statistical mechanics, and this means again that no entanglement is taken into account beyond short distances. The same is to be said about the product ansatz used in theories of quantum measurement and open quantum systems.  All this (and all the other features of condensed matter theory and statistical mechanics described in the previous sections) implies that there are limits to unitary time evolution and quantum superposition in macroscopic, finite-temperature systems. 

In fact, only as long as a quantum system does not interact with the rest of the world can it be described by the Schr\"odinger equation. When a system has finite temperature, it cannot avoid interaction with the rest of the world, as it emits thermal radiation. This is an irreversible process
that makes the time evolution of the entire system non-unitary.

Theories and interpretations of quantum mechanics that try to describe the interaction with a heat bath by a unitary time evolution cannot solve the puzzles and contradictions posed by quantum mechanics. Global application of unitary time evolution always leads to superpositions of states and histories (each of which might resemble classical states and histories), and not to a unique, stochastically chosen outcome as observed in experiments on quantum measurement. This is the criticism brought forward by several authors \cite{adler2003decoherence,schlosshauer2005decoherence} against the claim that decoherence solves the measurement problem or explains the quantum-classical transition. 

The most natural conclusion is therefore that there are limits of validity to the Schr\"odinger equation and to linear superposition. In fact, everybody would admit that the Schr\"odinger equation has limits of validity as it ignores relativistic effects, spin, and the coupling to the electrodynamic vacuum. Nevertheless, many scientists appear to think that these limitations of the Schr\"odinger equation do not become relevant in condensed-matter physics, in particular since there are established methods for including the spin degrees of freedom and transitions that involve the emission or absorption of photons. However, this is an unfounded belief, since all three properties of the Schr\"odinger equation (linear superposition, determinism, time symmetry) are violated by the methods employed in statistical mechanics and condensed matter theory, as this paper has explained. Such 'violations' are obviously needed in order to describe adequately the phenomena observed in condensed matter. 

This should remind us that  all our theories in physics are models of reality and not an exact image of reality. They are idealizations that neglect many features of a system in order to explain the phenomena or properties of interest. Therefore it should not be surprising that not everything can be described by one theory. Nevertheless, the challenge remains to figure out where the limits of the Schr\"odinger equation are in condensed matter systems. In the previous sections, we have pointed to the thermal wavelength and the thermal time. However, this applies only to those degrees of freedom that constitute the heat bath or that are in equilibrium with it. Superpositions, entanglement, and spatially extended quantum states can be sustained over considerable time in situations where the interaction with the thermal degrees of freedom is sufficiently weak. This applies to nuclear spins in NMR experiments and electron spins in ESR experiments, to photons that are entangled over many kilometers, or to macroscopic wave functions that constitute the superfluid or superconducting state and are protected by an energy gap from thermal excitations. 

In fact these macroscopic wave functions are not quantum states in the full sense. As mentioned above, they have both classical and quantum features. Their phase is a quantum feature, but the amplitude is proportional to the square root of the particle density, which means that states with different particle numbers are not orthogonal to each other. But they should be according to the Fock-space formalism. Furthermore, calculations that involve these macroscopic wave functions are often nonlinear in the wave function, as for instance the Ginzburg-Landau theory of superconductivity, or the Gross-Pitaevskii equation for Bose condensates. 

Crystals are another good example for classical features of a 'ground state' of a macroscopic system. They are the 'vacuum' for the phonons, just as a superfluid is a vacuum for rotons and phonons, and the BSC state is a vacuum for the Bogoliubov quasiparticles (which are unpaired electrons). Condensed matter physics thus shows us over and again that quantum systems depend on their classical context: First, the existence and properties of the considered 'vacuum', or medium, in which (quasi)particles live, is itself dependent on classical environmental variables such as temperature and pressure, and it assumes itself classical features. Second, the (quasi)particles themselves depend on the specific medium in which they occur. In view of this strong dependence of quantum systems on the classical context, a reductionist view that takes quantum mechanics as the foundation of everything, is untenable. We physicists need to learn to look at top-down influences in addition to bottom-up influences. 

Biological systems are a particularly intriguing example for the interaction between classical and quantum physics. It still is an open question over which spatial and temporal scales quantum excitations occur in biological systems. Since these systems are nonequilibrium systems with a complex structure, there are no a priori reasons why all these excitations should be confined to the thermal length and time scales. The emergent field of quantum biology will in the future certainly provide important insights into the limits of quantum superposition.

%warum diese Theorien nicht in der TOE enthalten sind. Die Physikphilosophen sollten sich dieses Themas genauer annehmen. Whin man schaut in CMT, sieht man diese Features.

%coupling to a heat bath sets limits to unitary time evolution, linear superposition, determinism, and time reversibility

%das paper von agata (PI) zur beschreibung dieser wellenpakete

%Heat baths do not follow unitary time evolution
%stochastic dynamics, this is inherited by the quantum system since it cannot become entangled with the heat bath

%There is a top-down influence from the macroscopic environment
%Classical system when in contact with a quantum system forces it into a localized state since entanglement is not possible

%is it clear how to describe a quantum system that is in contact with such a classical system? 

%what happens in non-equilibrium? Which part of the system is to be described classically, and what are the (local) quantum excitations? Quantum biology!

%\bibliographystyle{spbasic}
\bibliographystyle{unsrt}
%\bibliography{quantumlit_gesamt}
%

\end{document}